\documentclass[pra,twocolumn]{revtex4}

\usepackage{graphics}
\usepackage{graphicx}
\usepackage{bm}
\usepackage{latexsym}
\usepackage{color}
\usepackage{mathrsfs}
\usepackage{braket}
\usepackage{enumerate}
\usepackage[usenames,dvipsnames]{xcolor}
\usepackage[colorlinks=true,citecolor=blue,linkcolor=blue,urlcolor=blue,backref=true]{hyperref}
\usepackage{float}
\usepackage[raggedright,tight]{subfigure}  
\usepackage{amsmath}
\usepackage{amsfonts}
\usepackage{braket}
\usepackage[dvipsnames]{xcolor}
\usepackage{rotating}
\usepackage{multirow}
\usepackage{xcolor}

\begin{document}

\title{Gauge non-invariance due to material truncation in ultrastrong-coupling QED}
\author{Adam Stokes}
\author{Ahsan Nazir}
\address{Department of Physics and Astronomy, The University of Manchester, Oxford Road, Manchester M13 9PL, United Kingdom}

\date{\today}

\begin{abstract}
Gauge non-invariance due to material truncation has recently been explored in a number of contexts in strong-coupling QED. We show that the approach proposed recently in Nature Physics \textbf{15}, 803 (2019) rests on an incorrect mathematical assertion and so does not resolve gauge non-invariance. It produces new two-level models that are not equivalent in different gauges. The new Coulomb-gauge model is inaccurate for the regimes considered in Nature Physics \textbf{15}, 803 (2019), for which the multipolar-gauge quantum Rabi model is accurate. The models analysed in Nature Physics \textbf{15}, 803 (2019) do not result from the argument provided in the main text, but instead from truncation within the multipolar-gauge followed by the application of a truncated phase-invariance principle. More generally, this principle can be applied following truncation in any gauge and it yields an equivalence class of truncated models within the gauge chosen. Equivalence classes belonging to different gauges are not equivalent and the truncated phase-invariance principle does not provide an argument to prefer a particular class. In general, the optimal class depends on the physical situation, including the observables, the parameter regime, and the number of field modes being considered. We also emphasise that gauge-ambiguities are not synonymous with gauge non-invariance due to approximations. Independent of material truncation subsystem predictions can be vastly different depending on the theoretical definitions of the subsystems, which are controlled by the gauge choice. This constitutes an example of vector-space relativity that in no way contradicts gauge-invariance. However, within ultrastrong-coupling regimes this relativity can no longer be ignored.
\end{abstract}

\maketitle

\section*{Introduction}

Gauge-freedom in ultrastrong and deep-strong coupling QED has recently been investigated in a number of contexts \cite{stokes_gauge_2019,stokes_ultrastrong_2019,stokes_uniqueness_2019,stefano_resolution_2019,roth_optimal_2019,de_bernardis_breakdown_2018}. Its importance lies in the fact that 
light and matter quantum subsystems can only be defined relative to a gauge-frame in Hilbert space. An example implication, is that the predicted number of counts recorded by a photo-detector in an experiment will depend on the theoretical definition of photon that the detector is assumed to register. Each definition is gauge-invariant, but there are many possible gauge-invariant definitions and each one is provided by a different gauge. The Coulomb-gauge defines photons using the gauge-invariant field ${\bf E}_{\rm T}$ whereas the multipolar-gauge defines photons using the gauge-invariant field ${\bf D}_{\rm T}$. Arguments that a particular definition is universally correct are simplistic. It is far from clear, for example, that any conceivable photodetection device necessarily responds only to ${\bf E}_{\rm T}$-type photons. The differences between the available subsystem definitions can be subtle, being closely related to virtual particles and bare-energy conservation. This becomes much more important in ultrastrong-coupling regimes, wherein subsystem predictions will generally depend significantly on the adopted definitions.

One of the {\em many} implications of QED's subsystem gauge-relativity, is that approximations performed on a subsystem may ruin the gauge-invariance of the theory. An example is the truncation of the material subsystem to a finite number of energy levels \cite{stokes_gauge_2019,stokes_ultrastrong_2019,stokes_uniqueness_2019,stefano_resolution_2019,roth_optimal_2019,de_bernardis_breakdown_2018}. A proposal has recently been put forward to resolve this gauge non-invariance by attempting to define unitary two-level model versions of gauge-transformations \cite{stefano_resolution_2019}. 

It is claimed that gauge non-invariance resulting from material truncation is resolved through a method of deriving two-level models, which in any gauge yields a model equivalent to the multipolar-gauge quantum Rabi model (QRM). We show that this method rests crucially on the incorrect mathematical assertion that $Pf(O)P=f(POP)$ for a non-linear function $f$, projection $P\neq I$ and operator $O$. As a result, the method actually introduces yet further gauge non-invariance of the type that it seeks to resolve. It defines an entirely new class of two-level models that are not equivalent to each other, nor to any of the standard two-level models. We consider the example of a highly anharmonic double-well dipole 
as in Ref. \cite{stefano_resolution_2019}. 
{We show} that in the Coulomb-gauge the new type of two-level model gives a very inaccurate approximation of the exact theory. Contrary to the claim of Ref.~\cite{stefano_resolution_2019} the model is not equivalent to the multipolar-gauge QRM, which is very accurate, and is instead qualitatively very similar to the standard Coulomb-gauge QRM in the regimes considered therein.

We briefly review the possibility of defining a phase-invariance principle within a truncated theory using the truncated position operator following the idea presented in Supplementary Note 1 of Ref.~\cite{stefano_resolution_2019}. We show that such a principle can be invoked within the truncated theory of any gauge and results in an equivalence class of models therein. The models analysed in Ref.~\cite{stefano_resolution_2019} follow from imposing the truncated phase-invariance principle after initial truncation within the multipolar-gauge. These models are not those implied by the argument in the main text in Ref.~\cite{stefano_resolution_2019} and the phase-invariance principle used to obtain them does not in itself provide an argument to prefer initial truncation within the multipolar-gauge.

In regimes of large anharmonicity it is known that the multipolar-gauge QRM usually provides the most accurate predictions across the full energy spectrum when considering a single radiation mode \cite{de_bernardis_breakdown_2018}. This is certainly not a proof that the multipolar-gauge QRM will provide the most accurate predictions for any observable property of any system in any parameter regime. Indeed, it can be shown easily using the simple counter-example of a material oscillator, that this is mathematically impossible \cite{stokes_gauge_2019}. The optimal two-level model will depend upon the observable property being considered, as well as the parameter regime considered, and on the number of field modes included. Furthermore, in many situations a two-level material truncation is straightforwardly avoidable. Gauge-ambiguities are much broader than the prosaic gauge non-invariance that results from an approximation. In ultrastrong-coupling regimes, subsystem predictions vary significantly with the gauge relative to which the subsystems are defined, independent of model approximations. 

We remark at the outset, that the merit of a scientific work is subjective and it is not our intention to offer any opinions regarding merit. We attempt to confine our analysis to the mathematical facts and their immediate logical implications.

\section*{Gauge-freedom}

A single electron atom has Hamiltonian $H={\bf p}^2/(2m) +V({\bf r})$ where $[r_i,p_j]=i\delta_{ij}$ and $V$ is the binding potential. The momentum ${\bf p}=-i\nabla$ acts on the wave-function $\psi({\bf r})$. The gauge-principle asserts that the interaction of the atom with the electromagnetic field should be invariant under a local phase transformation $\psi'({\bf r}) = e^{iq\chi({\bf r})}\psi({\bf r})$ where $q$ is the electronic charge. This is guaranteed by making the minimal-coupling replacement $-i\nabla-q{\bf A}$ where any two gauge-fields ${\bf A}$ and ${\bf A}'={\bf A}+\nabla\chi$ are physically equivalent. According to the gauge-principle, the freedom to transform between distinct minimal-coupling prescriptions ${\bf p}-q{\bf A}$ and ${\bf p}-q{\bf A}'$ defines gauge-freedom.

Since $\nabla\chi$ is a longitudinal vector field, gauge-freedom is the freedom to choose ${\bf A}_{\rm L}$. A specific choice is accompanied by a specific local phase for $\psi$. In other words, gauge-freedom is the freedom to choose between different pairs $(\psi,{\bf A})$ and $(\psi',{\bf A}')$. 
%
It is clear that in order to implement ${\bf p} -q{\bf A} \to {\bf p} -q({\bf A}+\nabla \chi)$, the canonical momentum must transform as ${\bf p}\to e^{iq\chi({\bf r})}{\bf p}e^{-iq\chi({\bf r})} ={\bf p}- q\nabla \chi$. This transformation property relies upon the canonical operator algebra $[r_i,p_j]=i\delta_{ij}$, which cannot be supported by a finite-dimensional Hilbert space.

In a gauge transformation the vector potential changes by the gradient of a function, so the transverse vector potential ${\bf A}_{\rm T}$ is gauge-invariant. Gauge-freedom can be encoded into a real gauge-parameter $\alpha$ \cite{stokes_gauge_2019} as;
\begin{align}\label{A}
{\bf A}_\alpha({\bf x}) &= {\bf A}_{\rm T}({\bf x}) -\alpha \nabla \int_0^1 d\lambda\, {\bf x} \cdot {\bf A}_{\rm T}(\lambda {\bf x})\nonumber \\
&\equiv {\bf A}_{\rm T}({\bf x}) +\nabla \chi_\alpha({\bf x})
\end{align}
in which ${\bf A}_{\rm L}=\nabla\chi_\alpha$ is uniquely specified by choosing $\alpha$. Detailed derivations of the $\alpha$-gauge Hamiltonian $H_\alpha$ using Dirac's constrained quantisation procedure can be found in Refs. \cite{stokes_gauge_2019,stokes_uniqueness_2019}. This procedure reveals that when ${\bf A}_{\rm L}$ is a functional of ${\bf A}_{\rm T}$ as in Eq.~(\ref{A}), the choice of gauge $\alpha$ also uniquely specifies the arbitrary transverse polarisation defined by
\begin{align}
{\bf P}_{\rm T\alpha}({\bf x})=\alpha {\bf P}_{\rm T}({\bf x}) =\alpha \int_0^1 d\lambda\, q{\bf r}\cdot \delta^{\rm T}({\bf x}-\lambda{\bf r}).
\end{align}
{\em Unitary gauge-fixing transformations} $R_{\alpha\alpha'}$ are defined as
\cite{stokes_gauge_2019,lenz_quantum_1994,stokes_noncovariant_2012,chernyak_gauge_1995,stokes_uniqueness_2019}
\begin{align}\label{pzw}
R_{\alpha\alpha'} := \exp \left[i\int d^3 {\bf x} \, \left[{\bf P}_{\rm T\alpha}({\bf x})-{\bf P}_{\rm T\alpha'}({\bf x})\right] \cdot {\bf A}_{\rm T}({\bf x})\right].
\end{align}
Their {\em defining} property is to transform between distinct minimal-coupling prescriptions that appear within the Hamiltonian $H_\alpha$;
\begin{align}
&R_{\alpha\alpha'}\left[{\bf p}-q{\bf A}_\alpha({\bf r})\right]R_{\alpha\alpha'}^\dagger = {\bf p}-q{\bf A}_{\alpha'}({\bf r})\label{mina}\\
&R_{\alpha\alpha'}\left[{\bf \Pi}_{\rm T}+{\bf P}_{\rm T\alpha}\right]R_{\alpha\alpha'}^\dagger ={\bf \Pi}_{\rm T}+{\bf P}_{\rm T\alpha'}.\label{minb}
\end{align}
The effect of the transformation has been the replacement $({\bf A}_\alpha,{\bf P}_{\rm T\alpha}) \to ({\bf A}_{\alpha'},{\bf P}_{\rm T\alpha'})$, which clearly constitutes a gauge-transformation from the fixed gauge $\alpha$ to the fixed gauge $\alpha'$. As a result it follows that
\begin{align}\label{halp}
H_{\alpha'} = R_{\alpha\alpha'}H_{\alpha}R_{\alpha\alpha'}^\dagger.
\end{align}
This unitary equivalence of Hamiltonians of different gauges exemplifies the gauge-invariance of the theory.


\section*{Electric dipole and single mode approximations}

The electric-dipole approximation (EDA) and single-mode approximation can be performed preserving all algebraic properties of the theory, thereby preserving gauge-invariance \cite{stokes_gauge_2019,stokes_ultrastrong_2019,stokes_uniqueness_2019}. The dipole is assumed to be located at the origin ${\bf 0}$ and the canonical operators are assumed to point in the direction ${\bm \varepsilon}$ of polarisation of the single mode {of frequency $\omega$}. They are specified entirely by scalar operator components in this direction. Hereafter we use $x={\bm \varepsilon}\cdot {\bf r}$ to denote the dipole coordinate for ease of comparison with Ref.~\cite{stefano_resolution_2019}. We use $A={\bm \varepsilon}\cdot {\bf A}_{\rm T}$ to denote the component of the transverse vector potential, while $p$ and $\Pi$ are the corresponding dipole and cavity canonical momenta such that $[x,p]=i$ and $[A,\Pi]=i/v$ with $v$ the cavity volume.

The unitary gauge-fixing transformation between gauges $\alpha$ and $\alpha'$ is now given by \cite{stokes_gauge_2019,stokes_ultrastrong_2019}
\begin{align}\label{trans}
R_{\alpha\alpha'} = \exp (i [\alpha-\alpha']q xA).
\end{align}
Since gauge-fixing transformations remain unitary the gauge-invariance of the theory is preserved. The $\alpha$-gauge continues to be specified by its vector potential ${\bf A}_\alpha={\bm \varepsilon}A_\alpha$ and polarisation ${\bf P}_{\rm T\alpha}={\bm \varepsilon}P_{\rm T\alpha}$ which now read
\begin{align}
&A_\alpha= (1-\alpha)A,\label{Aa}\\
&P_{{\rm T}\alpha}={\alpha qx\over v}.\label{pt}
\end{align}
The {\em definition} of gauge-freedom given by Eqs.~(\ref{mina}) and (\ref{minb}) now reads
\begin{align}
&R_{\alpha\alpha'}pR_{\alpha\alpha'}^\dagger = p-(\alpha-\alpha')qA,\label{min1c}\\
&R_{\alpha\alpha'}\Pi R_{\alpha\alpha'}^\dagger =\Pi-(\alpha-\alpha'){qx\over v}\label{min2c}.
\end{align}
Were Eqs. (\ref{min1c}) and (\ref{min2c}) not satisfied, then $R_{\alpha\alpha'}$ would not effect the replacement $(A_\alpha,P_{\rm T\alpha})\to (A_{\alpha'},P_{\rm T\alpha'})$, meaning it would not be a gauge-transformation from $\alpha$ to $\alpha'$. Note that since $Uf(O)U^\dagger=f(UOU^\dagger)$ for any unitary transformation $U$, suitably well-defined function $f$, and operator $O$, Eqs.~(\ref{min1c}) and (\ref{min2c}) are necessary and sufficient to define how arbitrary functions of $p$ and $\Pi$ transform under a gauge-transformation.

Fundamentally, the Hamiltonian for a closed system is the total energy, which in this case is the sum of material mechanical and transverse electromagnetic energies \cite{stokes_gauge_2019,stokes_ultrastrong_2019};
\begin{align}
&H_\alpha = {\cal H}_{m}(A_\alpha)+{\cal H}_{\rm ph,\alpha}\label{edaham} \\
&{\cal H}_{m}(A_\alpha):={1\over 2}m{\dot x}^2+V(x) = {1\over 2m}\left(p-qA_\alpha\right)^2+V(x),\label{mechen1} \\ 
&{\cal H}_{\rm ph,\alpha}:= {v\over 2}(E_{\rm T}^2+\omega A^2) = {v\over 2}\left[(\Pi+P_{\rm T\alpha})^2+\omega^2A^2\right]\label{photen1}
\end{align}
where ${\dot x}=-i[x,H_\alpha]$ and $E_{\rm T}=-{\dot A}_{\rm T}= i[A_{\rm T},H_\alpha]$. Hamiltonians of different gauges are unitarily related as in Eq.~(\ref{halp}).

The method of Ref.~\cite{stefano_resolution_2019} relies on further properties of the Hamiltonian in Eq.~(\ref{edaham}), which hold only due to simplifications. In particular, within the EDA the gauge-function $\chi_\alpha$ becomes
\begin{align}
\chi_\alpha({\bf r}) = -\alpha {\bf r}\cdot {\bf A}_{\rm T}({\bf 0}),
\end{align}
which gives $\nabla \chi_1({\bf r}) = -{\bf A}_{\rm T}({\bf 0})$, and so ${\bf p}-q{\bf A}_1({\bf r}) \approx {\bf p}$. Thus, letting $\alpha=1$ on the left-hand-side of Eq.~(\ref{mina}) we obtain $R_{1\alpha}{\bf p}R_{1\alpha}^\dagger 
={\bf p}-q{\bf A}_\alpha$ where ${\bf A}_\alpha := (1-\alpha){\bf A}_{\rm T}({\bf 0})$ is the dipole approximation of ${\bf A}_\alpha({\bf r})$. In other words, within (and only within) the EDA the $\alpha$-gauge mechanical momentum may be obtained from the canonical momentum ${\bf p}$ using $R_{1\alpha}$. 
Within the full 3-dimensional setting and without the EDA it is impossible to implement the minimal-coupling prescription by unitary transformation of ${\bf p}$, because for any differentiable function $\chi$ we have $e^{-iq\chi({\bf r})}{\bf p}e^{iq\chi({\bf r})} = {\bf p}+q\nabla \chi({\bf r})$. The gradient $\nabla \chi$ is a longitudinal field therefore we cannot have $\nabla \chi({\bf r}) =-{\bf A}({\bf r})$ for all ${\bf r}$, because the transverse part ${\bf A}_{\rm T}$ of ${\bf A}$ is necessarily non-vanishing. What is fundamental and completely general is the gauge-transformation $e^{iq\chi({\bf r})}[{\bf p}-q{\bf A}({\bf r})]e^{-iq\chi({\bf r})} = {\bf p}-q[{\bf A}+\nabla \chi({\bf r})]$, which yields the result $R_{1\alpha}{\bf p}R_{1\alpha}^\dagger 
={\bf p}-q{\bf A}_\alpha$ as an approximate special case in which we let ${\bf A}={\bf A}_1$, $\chi=\chi_\alpha-\chi_1$, and we perform the EDA. 

Turning our attention to the mechanical energy, Eq.~(\ref{mina}) implies that this energy can be expressed in different gauges via gauge-transformation as 
\begin{align}\label{mechg}
{\cal H}_{m}(A_{\alpha'})=R_{\alpha\alpha'}{\cal H}_{m}(A_\alpha)R_{\alpha\alpha'}^\dagger,
\end{align}
which is simply an expression of the gauge-principle within the $\alpha$-gauge framework. However, as noted above, within the approximations made $A_1 = (1-\alpha)A|_{\alpha=1}\equiv 0$ and therefore ${\cal H}_{m}(A_1)$ is actually independent of $A$. Thus, for $\alpha=1$, Eq.~(\ref{mechg}) has the appearance of a unitary transformation applied to the free material Hamiltonian; 
\begin{align}\label{halpm2}
{\cal H}_{m}(A_{\alpha'}) = R_{1\alpha'}H_m R_{1\alpha'}^\dagger
\end{align}
where we have used that due to the EDA
\begin{align}\label{bareen}
{\cal H}_{m}(A_1) = H_m={p^2\over 2m}+V(x).
\end{align}

Turning now to the transverse electromagnetic energy, ${\cal H}_{\rm ph,\alpha}$, from Eq.~(\ref{minb}) we have
\begin{align}\label{phg}
{\cal H}_{\rm ph,\alpha'}=R_{\alpha\alpha'}{\cal H}_{\rm ph,\alpha}R_{\alpha\alpha'}^\dagger
\end{align}
and in particular, 
\begin{align}\label{bareenph}
{\cal H}_{\rm ph,\alpha} = R_{0\alpha}H_{\rm ph}R_{0\alpha}^\dagger
\end{align}
where
\begin{align}
H_{\rm ph}={\cal H}_{\rm ph,0}={v\over 2}(\Pi^2+\omega^2A^2).
\end{align}
Combining Eqs.~(\ref{halpm2}) and (\ref{bareenph}) we see that within the simplified setting of a one-dimensional model in the EDA the Hamiltonian can be written in terms of unitary gauge-fixing transformations of the free Hamiltonians $H_m$ and $H_{\rm ph}$ as
\begin{align}\label{halp2}
H_\alpha = R_{1\alpha}H_m R_{1\alpha}^\dagger + R_{0\alpha}H_{\rm ph} R_{0\alpha}^\dagger.
\end{align}
This is an approximate special case of the more general fundamental expression
\begin{align}\label{halp3}
H_\alpha = R_{\alpha'\alpha}{\cal H}_{m}(A_{\alpha'}) R_{\alpha'\alpha}^\dagger + R_{\alpha''\alpha}{\cal H}_{\rm ph,\alpha''} R_{\alpha''\alpha}^\dagger,
\end{align}
which expresses that both ${\cal H}_{m,\alpha}(A)$ and ${\cal H}_{\rm ph,\alpha}$ can be obtained as gauge-transformations of their counterparts in arbitrary gauges $\alpha'$ and $\alpha''$ respectively. This follows immediately from the definition of gauge-transformation given by Eqs.~(\ref{mina}) and (\ref{minb}). Using the subgroup property $R_{\alpha\alpha'} = R_{\alpha\beta}R_{\beta\alpha'}$ one can also deduce Eq.~(\ref{halp}) from Eq.~(\ref{halp3}). The fundamental equation (\ref{halp3}) reduces to the approximate equation (\ref{halp2}) when we choose $\alpha'=1$ and $\alpha''=0$, and we make use of ${\cal H}_{\rm ph,0} =H_{\rm ph}$ and ${\cal H}_{m}(A_1)=H_m$. It is this final equality that holds only because of the approximations and simplifying assumptions made, without which the derivation of ${\cal H}_{m}(A_\alpha)$ via unitary transformation of $H_m$ is impossible.


\section*{Material truncation}

Since the canonical momentum $p$ represents a different physical observable for each different value of $\alpha$, the same is true of $H_m$. Therefore, projecting onto a finite number of eigenstates of $H_m$ is a gauge-dependent procedure. Eigenvalues of $H_m$ are denoted $\epsilon_n$. The projection $P$ onto the first two-levels $\ket{\epsilon_0}$, $\ket{\epsilon_1}$ of $H_m$ gives $P H_m P= \omega_m \sigma^+ \sigma^- +\epsilon_0$ and $P qxP =d\sigma^x$ where $\sigma^+ = \ket{\epsilon_1}\bra{\epsilon_0}$, $\sigma^- = \ket{\epsilon_0}\bra{\epsilon_1}$ and $\sigma^x=\sigma^+ +\sigma^-$. The first transition energy is denoted $\omega_m=\epsilon_1-\epsilon_0$, and the transition dipole moment $d=\bra{\epsilon_1}qx\ket{\epsilon_0}$ is assumed to be real. More generally, one can define $P$ as the projection onto any finite number of levels.

That truncation necessarily ruins gauge-invariance is shown in Ref.~\cite{stokes_gauge_2019} where a passive perspective of rotations within the operator algebra is adopted. Within the active perspective adopted so far, different gauges possess unitarily related Hamiltonians $H_\alpha$ as specified by Eq.~(\ref{halp}). These Hamiltonians are expressed in terms of the same canonical momenta $p=m{\dot x}+qA_\alpha$ and $\Pi={\dot A}-P_{\rm T\alpha}$ where $m{\dot x}=-im[x,H_\alpha]$ and ${\dot A}=-i[A,H_\alpha]$. {On the other hand,} in the passive perspective the Hamiltonian $H$ is uniquely specified as the total energy operator of the system but different gauges possess different canonical momenta $p_\alpha =m{\dot x}+qA_\alpha$ and $\Pi_\alpha ={\dot A}-P_{\rm T\alpha}$ where $m{\dot x}=-im[x,H]$ and ${\dot A}=-i[A,H]$ \cite{stokes_gauge_2019}. It is now the momenta belonging to different gauges that are related as $p_\alpha = R_{\alpha\alpha'}p_{\alpha'}R_{\alpha\alpha'}^\dagger$ and $\Pi_\alpha = R_{\alpha\alpha'}\Pi_{\alpha'}R_{\alpha\alpha'}^\dagger$. The Hamiltonian $H$ can be expressed in terms of the momenta belonging to any gauge. The passive and active perspectives of rotations are strictly equivalent. 

When adopting the passive perspective it is seen that each gauge defines a different bare material Hamiltonian $H_m^\alpha = p_\alpha^2/(2m)+V(r)$. These Hamiltonians are related by $H_m^\alpha = R_{\alpha\alpha'}H_m^{\alpha'}R_{\alpha\alpha'}^\dagger$, therefore the corresponding two-level projections $P^\alpha$ and $P^{\alpha'}$ also satisfy $P^\alpha = R_{\alpha\alpha'}P^{\alpha'}R_{\alpha\alpha'}^\dagger$. Similarly, the photonic Hamiltonians $H_{\rm ph}^\alpha = {v\over 2}(\Pi_\alpha^2+\omega^2 A^2)$ belonging to different gauges are related by $H_{\rm ph}^\alpha = R_{\alpha\alpha'}H_{\rm ph}^{\alpha'}R_{\alpha\alpha'}^\dagger$. Thus, the bare basis eigenstates of $h^\alpha= H_m^\alpha+H_{\rm ph}^\alpha$ and $h^{\alpha'} = H_m^{\alpha'}+H_{\rm ph}^{\alpha'}$ are related by $\ket{\epsilon_n^\alpha,k^\alpha}=R_{\alpha\alpha'}\ket{\epsilon_n^{\alpha'},k^{\alpha'}}$. It should be noted that the tensor-product structures on the left and right-hand-sides of this equality are not the same. For non-vanishing coupling $R_{\alpha\alpha'}$ does not have the form $R_{\alpha\alpha'}= U \otimes V$ with respect to the $\alpha'$-gauge tensor-product factorisation of the Hilbert space. 

Each gauge defines a distinct two-level subspace, $P^\alpha{\cal H}$, within the full Hilbert space ${\cal H}$. The identity in ${\cal H}$ may be resolved in any orthonormal basis;
\begin{align}
I=\sum_{n=0}^\infty\sum_j \ket{\epsilon_n^{\alpha},j^{\alpha}}\bra{\epsilon_n^{\alpha},j^{\alpha}} = \sum_{n=0}^\infty\sum_j \ket{\epsilon_n^{\alpha'},j^{\alpha'}}\bra{\epsilon_n^{\alpha'},j^{\alpha'}}
\end{align}
and the operator $R_{\alpha\alpha'}$ may be written
\begin{align}
R_{\alpha\alpha'} = \sum_{n=0}^\infty\sum_j \ket{\epsilon_n^{\alpha},j^{\alpha}}\bra{\epsilon_n^{\alpha'},j^{\alpha'}}.
\end{align}
We therefore see that an arbitrary separable state $\ket{\psi_2^\alpha} =\sum_{i=0,1}\sum_k \beta_i \gamma_k \ket{\epsilon^\alpha_i,k^\alpha}$ belonging to the $\alpha$-gauge {two-level} subspace $P^\alpha{\cal H}$ may be expanded in the basis $\{\ket{\epsilon_n^{\alpha'},k^{\alpha'}}\}$ as
\begin{align}
\ket{\psi^\alpha_2} &= \sum_{n=0}^\infty \sum_{i=0}^1 \sum_{j,k} \beta_i \gamma_k \bra{\epsilon_n^{\alpha'},j^{\alpha'}}R_{\alpha\alpha'} \ket{\epsilon_i^{\alpha'},k^{\alpha'}} \ket{\epsilon^{\alpha'}_n,j^{\alpha'}} \nonumber \\ &\not \in P^{\alpha'}{\cal H}.
\end{align}
Therefore, the subspace defined by material truncation is not gauge-invariant. The truncated state space within the gauge $\alpha$ comprises states that are superpositions of {\em all} of the material basis states $\{\ket{\epsilon^{\alpha'}_n}\}$ belonging to the gauge $\alpha'$. Furthermore, the state $\ket{\psi_2^\alpha}$, which is separable with respect to the $\alpha$-gauge {light and matter subsystem eigenstates}, is entangled with respect to the $\alpha'$-gauge {light and matter subsystem eigenstates}. A projection onto a finite number of material states in one gauge constitutes a non-trivial operation on {\em both} subsystems in any other gauge.

The projection of $\ket{\psi^\alpha_2}\in P^\alpha{\cal H}$ onto the subspace $P^{\alpha'}{\cal H}$ is $P^{\alpha'}\ket{\psi_2^\alpha} = P^{\alpha'}R_{\alpha\alpha'}P^{\alpha'}\ket{\psi_2^{\alpha'}}$, where $\ket{\psi_2^{\alpha'}} =\sum_{i=0,1}\sum_k \beta_i \gamma_k \ket{\epsilon^{\alpha'}_i,k^{\alpha'}}\in P^{\alpha'}{\cal H}$. Evidently, this defines a non-unitary two-level model gauge transformation, ${\cal G}_{\alpha\alpha'}=P^{\alpha'}R_{\alpha\alpha'}P^{\alpha'}$, on $P^{\alpha'}{\cal H}$, while the restricted projection $P^{\alpha'}|_{P^\alpha{\cal H}}:P^\alpha{\cal H}\to P^{\alpha'}{\cal H}$ defined as a map between distinct two-level subspaces is also clearly non-unitary. Conversely, by definition, a two-dimensional unitary operator (Bloch-sphere rotation) of the form ${\cal T} = \exp[iP^\alpha f(x,A)P^\alpha]$ where $f$ is an arbitrary Hermitian function of $x$ and $A$, is a map ${\cal T}:P^\alpha{\cal H}\to P^\alpha{\cal H}$ and so it cannot transform between the truncated spaces $P^\alpha{\cal H}$ and $P^{\alpha'}{\cal H}$ of different gauges. In this sense, material truncation necessarily breaks gauge-invariance, {and it is not possible to define a two-level model unitary 
of the form ${\cal T}$ that also constitutes a gauge transformation. Here, the term {\em gauge} possesses the definition that is provided by the fundamental non-truncated theory. This is the only relevant definition if our aim is to determine the effect of truncation on the gauge-invariance of the starting theory. We shall discuss these points in the context of Ref.~\cite{stefano_resolution_2019} below}.

\section*{Truncated Hamiltonians}

For ease of comparison with Ref.~\cite{stefano_resolution_2019} we return to the active perspective of unitary rotations. A two-level truncation of the Hamiltonian $H_\alpha$ is a $P$-dependent map $M_P: H_\alpha\to M_P(H_\alpha)$, such that $M_P(H_\alpha) : P{\cal H} \to P{\cal H}$ is an Hermitian operator on $P{\cal H}$. Clearly such maps are not unique. The map yielding what we refer to as the ``standard" $\alpha$-gauge two-level model, consists of replacing $x$ and $p$ with their projected counterparts $PxP$ and $P p P$ as follows
\begin{align}\label{stnd}
M_P(H_\alpha) = H_\alpha^2 = PH_mP+PH_{\rm ph}P +{\cal V}^\alpha(PxP,PpP)
\end{align}
where ${\cal V}^\alpha(x,p)=H_\alpha-H_m-H_{\rm ph}$ is the interaction Hamiltonian. For distinct values of $\alpha$ the Hamiltonians $H^2_\alpha $ are not equivalent to each other \cite{stokes_gauge_2019,de_bernardis_breakdown_2018,stokes_ultrastrong_2019}. This is because physically, $P$ represents a different projection in each different gauge, {as discussed above}.

The definition (\ref{stnd}) is noteworthy, because it is capable of yielding the standard QRM that is so often encountered in light-matter physics; $H_{\rm QRM} = \omega_m\sigma^+\sigma^- + \omega a^\dagger a + ig(\sigma^+ + \sigma^-)(a^\dagger -a)$, where the frequencies $\omega_m$ and $\omega$ are independent of the model's coupling $g$. Specifically, a model of this form is obtained up to a zero-point shift by using Eq.~(\ref{stnd}) and choosing $\alpha=1$. For $\alpha\neq 1$ the Hamiltonian contains a coupling-dependent $A^2$-term such that Eq.~(\ref{stnd}) yields a bilinear coupling (as in $H_{\rm QRM}$) only if this term is included as part of $H_{\rm ph}$ rather than as part of ${\cal V}^\alpha$. If this is done, then the cavity frequency is coupling-dependent. Thus, for $\alpha\neq 1$ the model resulting from Eq.~(\ref{stnd}) cannot strictly coincide with $H_{\rm QRM}$.

Many alternatives to the particular definition of $M_P$ in Eq.~(\ref{stnd}) are of course available. An obvious alternative, $M_P'$, is straightforwardly defined by $M_P'(H_\alpha)=P H_\alpha P$. For $\alpha=0$ the map $M_P'$ coincides with
$M_P$ in Eq.~(\ref{stnd}), and as already noted this does not yield a strictly standard QRM. For $\alpha\neq 0$ the two maps differ, such that unlike $M_P$ the map $M_P'$ does not yield a strictly standard QRM even for $\alpha=1$. This is because ${\cal V}^\alpha$ contains a term quadratic in $x$. Resolving the identity as $I=P+Q$ we have $Px^2P = Px(P+Q)xP$ implying that, for $\alpha\neq 0$, the definition $M_P'(H_\alpha)$ only coincides with $M_P(H_\alpha)$ in Eq.~(\ref{stnd}) if $PxQ$ and $QxP$ can be neglected. For sufficiently anharmonic material systems this will be the case, but for less anharmonic material systems the difference between the two definitions may be more significant.

A third definition of two-level model results from including the term quadratic in $x$ within $H_m$ rather than ${\cal V}^\alpha$. This then results in an explicitly $\alpha$ and coupling-dependent projection ${\tilde P}^\alpha$ (which should not be confused with $P^\alpha$ defined in the previous section). Using this projection one may define the two-level model ${\tilde M}_P(H_\alpha)={\tilde P}^\alpha H_\alpha {\tilde P}^\alpha$, which is in general different to both $M_P(H_\alpha)$ and $M_P'(H_\alpha )$. A fourth method of defining two-level models is presented in Ref.~\cite{stefano_resolution_2019} and is reviewed below.

We note that all of the examples of truncating maps listed above result in models that are not equivalent in different gauges. We do not intend here to advocate any one map $M_P$ as being the most ``correct". We merely point out that different maps exist. 
What we are concerned with, is whether or not the procedures given in Ref.~\cite{stefano_resolution_2019} can be claimed to ``resolve" gauge non-invariance due to truncation. In the main text, Ref.~\cite{stefano_resolution_2019} seeks to derive two-level models from the free theory by replacing the unitary transformation $R_{\alpha\alpha'}$ in Eq. (\ref{halp2}) with a two-level model counterpart, which we will denote by ${\cal G}_{\alpha\alpha'}$. However, Ref.~\cite{stefano_resolution_2019} tacitly equates what are actually two quite different two-level model versions of $R_{\alpha\alpha'}$. This stems from the inequality
\begin{align}\label{lem2}
Pf(x)P \neq f(PxP)
\end{align}
holding for any non-linear function $f$. We have already noted an example of this inequality, namely, the case $f(x)=x^2$. We saw that this example implied that $M_P\neq M_P'$ for $\alpha\neq 0$. We reiterate here that our aim is not to advocate any particular map $M_P:H_\alpha\to M_P(H_\alpha)$ as being ``correct" because of the way it maps the function $f(x)=x^2$ within the Hamiltonian. Rather, what we are concerned with is the separate question of whether the left- and right-hand-sides of inequality (\ref{lem2}) can be said to be equal when $f(x)=R_{\alpha\alpha'}$. It is this question that is crucial to determining whether or not the argumentation presented in Ref.~\cite{stefano_resolution_2019} is sound.

In order to be absolutely precise regarding the meaning of inequality~(\ref{lem2}) we first establish definitions and notation. The Hilbert space of the light-matter theory is denoted ${\cal H}$. The projection operator $P$ is a map $P:{\cal H}\to P{\cal H}$ where $P{\cal H}:=\{\ket{\psi^2}=P\ket{\psi}:\ket{\psi}\in {\cal H}\}\subset {\cal H}$. The identity operator $I:{\cal H}\to {\cal H}$ can be partitioned as $I=P+Q$ where $I-P=Q:{\cal H}\to Q{\cal H}$ with ${\cal H}=P{\cal H}\oplus Q{\cal H}$. The identity operator on the space $P{\cal H}$ is denoted $I_2:P{\cal H}\to P{\cal H}$. It satisfies $I_2=P|_{P{\cal H}}$ where $P|_{P{\cal H}}$ denotes the restriction of $P$ to the domain $P{\cal H}$. The position operator $x$ is a map $x:{\cal H}\to {\cal H}$ whose range is the entire codomain; $x({\cal H}) = {\cal H}$. 

Two operators $A:V\to V$ and $B:V\to V$ defined on linear space $V$ are said to be equal if $A\ket{\psi}=B\ket{\psi}$ for all $\ket{\psi}\in V$. Since operators are maps on $V$, the notion of equality within the operator space ${\cal L}(V)$ is inherited from the notion of equality of vectors in $V$. Let $\circ$ denote composition of maps. The pair $({\cal L}(V),\circ)$ is an associative algebra. The algebraic product of $A$ and $B$ written as the juxtaposition $AB$ denotes the composition $A\circ B$. 

By definition of $x$ the juxtaposition $I_2 x$ is ill-defined on any nontrivial subspace within ${\cal H}$. In particular, letting $\ket{\psi_2}\in P{\cal H}$ we have immediately that the vector $x\ket{\psi_2}$ is not an element of $P{\cal H}$ (the domain of $I_2$) so  $I_2 x$ is not well-defined on $P{\cal H}$. The juxtaposition $Px$ is well-defined because the domain of $P$ is the range of $x$. The operator $P$ is not an identity operator unless restricted to $P{\cal H}$, but if $P$ is restricted to $P{\cal H}$ then its algebraic right multiplication by $x$ is not well-defined. It follows that $P=P^2$ cannot be inserted in between factors of $x$ within $x^n$ and simultaneously be treated as an identity operator. The operators $Px^nP$ and $(PxP)^n$ both have support and range $P{\cal H}$ and both are well-defined, but they are not {\em equal}. Inequality (\ref{lem2}) then follows from assuming that $f$ admits a series expansion in $x$.

We remark that inequality (\ref{lem2}) is not subject to interpretation. There is no freedom to interpret $P$ in different ways. All relevant operators $P$, $x$, and $I_2$ are uniquely defined. In particular, $P$ and $I_2$ are not different interpretations of the same operator, they are different operators, because they have different domains. Only $Px$ is well-defined whereas $I_2x$ is not. 
If $P$ truncates, then $P$ is not, and does not act as, an identity operator within the composition $Px$. 
The only circumstance under which inequality (\ref{lem2}) becomes an equality for a non-linear function $f$ is the case in which $P$ is the identity on the range of $x$, in which case $P$ does not perform a truncation. It performs no operation at all and inequality (\ref{lem2}) simply becomes $f(x)=f(x)$.

Due to inequality (\ref{lem2}) there are two different two-level model variants of the PZW transformation $R_{10}$, which are
\begin{align}\label{r10}
&{\cal G}_{10} = PR_{10}P = P\exp[iqxA]P, \\
&{\cal T}_{10} =  \exp[iqPxPA]\neq {\cal G}_{10}.\label{tcal}
\end{align}
Ref.~\cite{stefano_resolution_2019} claims that replacing $R_{10}$ with ${\cal G}_{10}$ in Eq.~(\ref{halp2}), which holds only due to the EDA, gives the correct Coulomb-gauge two-level model
\begin{align}\label{cg0}
{\tilde H}_0^2 = {\cal G}_{10}PH_mP{\cal G}_{10}^\dagger + PH_{\rm ph}P.
\end{align}
This derivation is described as constituting the ``correct application of the gauge principle", the idea being that non-localities introduced by the projection of the atomic potential contained in $H_m$ have now been properly accounted for. In passing between equations (8) and (9) of Ref.~\cite{stefano_resolution_2019} it is tacitly and incorrectly assumed that ${\cal G}_{01}={\cal T}_{01}$ from which it would follow that ${\tilde H}_0^2 = h_1^2(0)$ where
\begin{align}\label{cg2t}
h_1^2(0) = {\cal T}_{10}PH_mP{\cal T}_{10}^\dagger + PH_{\rm ph}P.
\end{align}
However, due to inequality (\ref{lem2}) the two-level models ${\tilde H}_0^2$ and $h_1^2(0)$ are actually very different. As {previously discussed and shown explicitly} below, $h_1^2(0)$ is not a Coulomb-gauge two-level model, because ${\cal T}_{10}$ cannot implement a gauge-change. Rather $h_1^2(0)$ is simply a Bloch-sphere rotation of the multipolar-gauge QRM. In contrast, the model ${\tilde H}_0^2$ is a Coulomb-gauge two-level model because ${\cal G}_{10}$ does implement a gauge change. However, ${\cal G}_{10}$ is not unitary and {so it does not preserve gauge invariance. In particular} ${\tilde H}_0^2$ is not equivalent to the multipolar-gauge QRM. 

\section*{Generalisation}

Generalising Eq.~(\ref{r10}), we see that there are two different two-level model versions of $R_{\alpha\alpha'}$ defined as
\begin{align}
&{\cal G}_{\alpha\alpha'} = PR_{\alpha\alpha'}P = P\exp[iq(\alpha-\alpha')xA]P \\
&{\cal T}_{\alpha\alpha'} =  \exp[iq(\alpha-\alpha')PxPA]\neq {\cal G}_{\alpha\alpha'}.\label{tcal}
\end{align}
The first of these transformations, ${\cal G}_{\alpha\alpha'}$, is not unitary contrary to the claim of Ref.~\cite{stefano_resolution_2019} that ${\cal G}_{10}$ is unitary. However, as noted in Ref.~\cite{stefano_resolution_2019}, ${\cal G}_{10}$ does implement a gauge change as defined by Eq.~(\ref{min1c}). More generally, if we let $POP=f(p)$, where $O$ is arbitrary, then
\begin{align}\label{projmin}
{\cal G}_{\alpha\alpha'}f(p) {\cal G}_{\alpha\alpha'} = Pf(p-(\alpha-\alpha')qA)P.
\end{align}
This follows immediately from Eq. (\ref{min1c}). In words, ${\cal G}_{\alpha\alpha'}$ implements a gauge-transformation within a projected operator and then re-projects the result. This property is crucial to the validity of the argumentation of Ref.~\cite{stefano_resolution_2019} which proposes that to properly account for non-localities introduced by the truncation, the minimal-coupling prescription $p\to p-qA$ should be implemented within $f(p)=PH_mP$ and the result re-projected. By this argument, replacing $R_{\alpha\alpha'}$ in Eq.~(\ref{halp2}) with ${\cal G}_{\alpha\alpha'}$, one obtains a new kind of (``correct") two-level model
\begin{align}\label{halpP}
{\tilde H}_\alpha^2 =& {\cal G}_{1\alpha}PH_mP{\cal G}_{1\alpha}^\dagger+  {\cal G}_{0\alpha}PH_{\rm ph}P {\cal G}_{0\alpha}^\dagger.
\end{align}
These models are not equivalent for different $\alpha$. Thus, the method claimed to be correct in Ref.~\cite{stefano_resolution_2019} is incapable of resolving the gauge non-invariance incurred by truncation. Indeed, such non-invariance is a necessary implication of truncation. Nevertheless, as already noted, the $\alpha=0$ model ${\tilde H}_0^2$ is claimed to be the ``correct" Coulomb-gauge two-level model in Ref. \cite{stefano_resolution_2019}.

The other two-level model transformation ${\cal T}_{\alpha\alpha'}$ which is given in Eq. (\ref{tcal}) is clearly unitary (unlike ${\cal G}_{\alpha\alpha'}$), but it does not implement a gauge change;
\begin{align}\label{nprojmin}
{\cal T}_{\alpha\alpha'}f(p) {\cal T}_{\alpha\alpha'} \neq Pf(p-(\alpha-\alpha')qA)P.
\end{align}
Indeed, a two-level model unitary transformation {\em cannot} implement the minimal-coupling replacement $p\to p-qA$ that is fundamental to the gauge-principle. By replacing $R_{\alpha\alpha'}$ in Eq.~(\ref{halp2}) with ${\cal T}_{\alpha\alpha'}$ one obtains two-level models
\begin{align}\label{halpP}
h_1^2(\alpha) =& {\cal T}_{1\alpha}PH_mP{\cal T}_{1\alpha}^\dagger+  {\cal T}_{0\alpha}PH_{\rm ph}P {\cal T}_{0\alpha}^\dagger \nonumber \\ =&{\cal T}_{1\alpha}H_1^2{\cal T}_{1\alpha}^\dagger
\end{align}
where the second equality shows that these models are equivalent to the standard multipolar-gauge QRM $H_1^2$. In particular, $h_1^2(1)=H_1^2$, because truncation does not alter the cavity canonical operator algebra. Since ${\cal T}_{1\alpha}$ does not implement a gauge change the parameter $\alpha$ in $h_1^2(\alpha)$ does not select a gauge. In this sense, one could equally well generate equivalent two-level models to $H_1^2$ using any two-level model unitary rotation whatsoever.

An equivalence class of models can be defined for each individual gauge $\alpha$ as $h_\alpha^2(\alpha') = {\cal T}_{\alpha\alpha'}H_\alpha^2{\cal T}_{\alpha\alpha'}^\dagger$. It will be shown later that such an equivalence class can be obtained via the imposition of a form of truncated phase-invariance within the $\alpha$-gauge truncated Hilbert space. For now we note that the standard $\alpha$-gauge model $H_\alpha^2 = h_\alpha^2(\alpha)$ is a representative from its class $\{h_\alpha^2(\alpha')\}$. In the notation $h_\alpha^2(\alpha')$ the subscript $\alpha$ denotes the gauge within which the truncation has been made, while the argument $\alpha'$ parameterises a position on the Bloch-sphere within the gauge $\alpha$. Gauge non-invariance cannot be resolved through the mere construction of models equivalent to some chosen two-level model. For any two distinct {\em gauges}, $\alpha_1$ and $\alpha_2$ within which truncation has been performed, the distinct equivalence classes $\{h_{\alpha_1}^2(\alpha')\}$ and $\{h_{\alpha_2}^2(\alpha')\}$ are not equivalent.

The inequality ${\cal G}_{\alpha\alpha'} \neq {\cal T}_{\alpha\alpha'}$ is not subject to interpretation. Neither is it an inconsequential technicality with regards to the argument presented in the main text of Ref.~\cite{stefano_resolution_2019}. The argument given is that the transformation ${\cal G}_{10}$ produces the correct Coulomb-gauge two-level model because unlike $PR_{10}$ the operator ${\cal G}_{10}=PR_{10}P$ performs the minimal-coupling replacement within the additional $p$-dependent terms that result from prior projection by $P$. Eqs.~(\ref{projmin}) and (\ref{nprojmin}) express the fact that ${\cal G}_{10}$ has this property whereas ${\cal T}_{10}$ (like $PR_{10}$) does not. Thus, only the models  ${\tilde H}_\alpha^2$ result from the argument that is provided in the main text of Ref.~\cite{stefano_resolution_2019} and these models are not equivalent to each other in different gauges. The models actually analysed, $\{h_1^2(\alpha)\}$, do not result from the procedure described in the main text.


The question that now naturally arises is whether or not the new Coulomb-gauge two-level model ${\tilde H}_0^2$ is {\em approximately} equivalent 
{to the} multipolar-gauge models $h_1^2(\alpha)$. In other words, can inequality (\ref{lem2}) ever become an approximate equality in the case that  $f(x)=R_{1\alpha}$? We have already noted one instance in which inequality (\ref{lem2}) does indeed become an approximate equality, this being the example of $f(x)=x^2$ under the condition that the contributions $PxQ$ and $QxP$ are negligible. We also noted that we might reasonably expect this latter condition to hold for sufficiently anharmonic material systems. 

We now apply the same analysis in the case that $f(x)=R_{10}$. 
Letting $I=P+Q$, we express the left-hand-side of inequality (\ref{lem2}), which is $PR_{10}P={\cal G}_{10}$, in the form
\begin{align}\label{ap1}
&{\cal G}_{10}\nonumber \\ &=P \exp\left[i\eta(\sigma^x+[PxQ+QxP+QxQ]/{\bar x})(a^\dagger +a)\right]P
\end{align}
where $\eta=d/\sqrt{2\omega v}$ is a dimensionless coupling parameter, ${\bar x}=\bra{\epsilon_0}x\ket{\epsilon_1}=d/q$ and $\sigma^x = PxP/{\bar x}$. If we assume that $PxQ \ll PxP$ and based on this we assume that the terms $PxQ$ and $QxP$ can be neglected in the exponent of $R_{10}$ then we obtain
\begin{align}\label{ap2}
{\cal G}_{10}&\approx P \exp\left[i\eta(\sigma^x+QxQ/{\bar x})(a^\dagger +a)\right]P \nonumber \\ &= P \exp\left[i\eta\sigma^x(a^\dagger +a)\right]P ={\cal T}_{10}
\end{align}
where we have used $PQ=0=QP$. One might suppose that such an approximation can be justified for a sufficiently anharmonic material system. However, if this is the case, then by following exactly the same steps one obtains $PR_{\alpha\alpha'}\approx {\cal T}_{\alpha\alpha'}$. From this one obtains $H_0^2\approx h_1^2(0)$ where the left-hand-side denotes the {\em standard} Coulomb-gauge Rabi model [given by Eq.~(\ref{stnd})] and the right-hand-side is equivalent to the standard multipolar-gauge Rabi model $H_1^2$. Since it is known that $H_0^2$ and $H_1^2$ are markedly different \cite{stefano_resolution_2019,stokes_gauge_2019,de_bernardis_breakdown_2018}, it follows that in general, one cannot neglect terms $PxQ$ and $QxP$ in the exponent of $R_{10}$ even for highly anharmonic material systems. In the following section it will be verified by an explicit example that the new Coulomb-gauge Hamiltonian ${\tilde H}_0^2$ is very different to the multipolar-gauge models $h_1^2(\alpha)$. {Thus,} contrary to the claim of Ref.~\cite{stefano_resolution_2019}, 
{${\tilde H}_0^2$} is not equivalent to the multipolar-gauge models $h_1^2(\alpha)$, among which is the standard multipolar-gauge QRM $H_1^2$. 

Inequality (\ref{lem2}) does become an approximate equality if it is permissible to expand the function $f$ to no higher than linear order in $x$. The exponent of $R_{\alpha\alpha'}$ is linear in $qx$, therefore the approximate equality ${\cal T}_{\alpha\alpha'} \approx {\cal G}_{\alpha\alpha'}$ does result if the exponentials on both sides can be expanded to linear order in $q$. In this case the two-level models ${\tilde H}_\alpha^2$ are then the same as the models $h_1^2(\alpha)$ and they must be equivalent to each other for different $\alpha$. However, a first order expansion of the model $h_1^2(\alpha)$ simply gives back the standard two-level model $H_\alpha^2$ to linear order in $q$. It follows that to this order, i.e., in the weak {light-matter} coupling regime, {\em all} two-level models are the same ${\tilde H}_\alpha^2=h_1^2(\alpha)=H_\alpha^2$. This is the only regime in which such an equivalence can generally be obtained.

It is incorrect to assume that this implies that the standard Coulomb-gauge QRM $H_0^2$ is somehow a first order approximation of the multipolar gauge QRM $H_1^2$. To see this note that analogous to the multipolar two-level models $h_1^2(\alpha)$, one can construct an equivalence class of Coulomb-gauge two-level models as $h_0^2(\alpha)={\cal T}_{0\alpha}H_0^2{\cal T}_{0\alpha}^\dagger$. To linear order in $q$ the models $h_0^2(\alpha)$ and $H_\alpha^2$ also coincide. In particular, if one expands up to first order in $q$ the $\alpha=1$ member of the Coulomb-gauge class, namely $h_0^2(1)$, then one obtains $H_1^2$ to linear order in $q$. Therefore, the multipolar-gauge QRM could equally well be viewed as a first order approximation of the standard Coulomb-gauge QRM.

As the coupling-strength increases the first order expansion in $q$ becomes progressively worse, so ${\cal T}_{\alpha\alpha'}$ and ${\cal G}_{\alpha\alpha'}$ become progressively different. Thus, if a particular model $H_{\alpha_1}^2$ were found to be accurate for some particular observable in some particular situation, then as the coupling increases any other model $H^2_{\alpha_2},~\alpha_2\neq \alpha_1$ will become progressively less accurate by comparison. As we have already noted, which of the standard models $H_\alpha^2$ is indeed most accurate, depends on the observable, the parameter regime, and the number of field modes being considered. For a single-mode and highly anharmonic matter the multipolar model $H_1^2$ tends to be more accurate across the full energy spectrum and in particular for higher levels \cite{stokes_gauge_2019,stefano_resolution_2019,roth_optimal_2019,de_bernardis_breakdown_2018}. For lower anharmonicity $H_1^2$ is not always optimal for the lowest two levels even for a single mode \cite{stokes_gauge_2019}. When considering more than one mode the multipolar-gauge QRM may be suboptimal even in regimes where two-level models can be generally accurate \cite{roth_optimal_2019}. It is easy to show that the form of the new Coulomb-gauge model ${\tilde H}_0^2$ depends strongly on the chosen dipolar potential, and that it only coincides with $h_1^2(0)$ to first order in $q$. To this order it also coincides with the standard Coulomb-gauge QRM $H_0^2$ as expected.

\section*{Gauge ambiguities}

Gauge non-invariance and gauge-ambiguities are not synonymous because a straightforwardly avoidable approximation cannot result in genuine ambiguities. The latter instead result from there existing several answers to a single question, such as the amount of light-matter entanglement in the ground state of an ultrastrongly coupled light-matter system. Independent of material truncation and the associated gauge non-invariance, ambiguities can occur in the description of, for example, time-dependent interactions \cite{stokes_ultrastrong_2019}, and Dicke-model superradiance \cite{stokes_uniqueness_2019}. Subsystem predictions such as photon population depend on the gauge relative to which the subsystems are defined, with important implications. For example, the gauge controls the classification of the Dicke-model quantum phase transition as being radiative versus non-radiative \cite{stokes_uniqueness_2019}. 

Uniquely predicting the number of counts recorded by some specific photodetection device in an experiment requires us to specify within the theory, which photons the device registers. Each gauge provides us with a physically different gauge-invariant definition of ``photon". What varies between these definitions are the spacetime localisation properties of ``material sources". This facet is intimately related to the inclusion or exclusion of virtual quanta within the definition of a ``material source". Measurements possess a certain extent in both space and time, and the most operationally relevant definition of ``photon" may well depend on the time and length-scales of the measurements available.

Stipulating that one or another definition of a quantum subsystem is universally correct (i.e., relevant) would seem to be much too simplistic. For example, the Coulomb-gauge definition of photon uses the transverse electric field ${\bf E}_{\rm T}={\bf E}-{\bf E}_{\rm L}$, which is a non-local field, because ${\bf E}$ is local by fundamental assumption while the electrostatic field ${\bf E}_{\rm L}$ is non-local, being instantaneously connected to its source. The Coulomb-gauge conception of a ``material source" which generates the field ${\bf E}_{\rm T}$, implicitly includes not only purely mechanical degrees of freedom but also all longitudinal-electric degrees of freedom as well. In particular, the Coulomb-gauge material momentum of a dynamical charge $q$ at ${\bf r}$ is
\begin{align}\label{coul0}
{\bf p} =m{\dot {\bf r}}+q{\bf A}_0({\bf r}) = m{\dot {\bf r}} + \int d^3 x\, {\bf E}_{\rm L,{\bf r}}({\bf x})\times {\bf B}({\bf x})
\end{align}
where $q{\bf A}_0({\bf r})=q{\bf A}_{\rm T}({\bf r})= \int d^3 x\, {\bf E}_{\rm L,{\bf r}}({\bf x})\times {\bf B}({\bf x})$ is the momentum associated with the static field ${\bf E}_{\rm L,{\bf r}}$ generated by the bare charge; $\nabla \cdot {\bf E}_{\rm L,{\bf r}}({\bf x})=q\delta({\bf x}-{\bf r})$. Thus, in the Coulomb-gauge ${\bf p}$ is clearly a gauge-invariant observable, but it is obviously not purely mechanical. The corresponding energy of the ``material" charge is
\begin{align}
H_m={{\bf p}^2\over 2m} + V({\bf r})
\end{align}
where we have allowed an arbitrary potential. This energy is similarly not purely mechanical. The Coulomb-gauge conceptions of ``matter" and ``light" are ones in which ``matter" is not fully localised, it includes all electrostatic contributions, meaning that these degrees of freedom are absent from ``light", which is defined in terms of ${\bf E}_{\rm T}$ alone.

Unlike in the Coulomb-gauge, in the multipolar-gauge 
\begin{align}\label{mult0}
{\bf p}=m{\dot {\bf r}}+ q{\bf A}_1({\bf r}) = m{\dot {\bf r}}-q\int_0^1 d\lambda\, \lambda{\bf r} \times {\bf B}(\lambda{\bf r})
\end{align}
which is also clearly a gauge-invariant observable, but obviously a different observable to that in Eq.~(\ref{coul0}). The multipolar potential ${\bf A}_1({\bf r}) = -\int_0^1 d\lambda\, \lambda {\bf r}\times {\bf B}(\lambda{\bf r})$ is clearly more localised than ${\bf A}_0({\bf r})$ in Eq.~(\ref{coul0}). It depends on the local field ${\bf B}$ at points on the straight-line connecting the multipolar-centre at ${\bf 0}$ to the charge at ${\bf r}$. This domain is by definition {\em inside} the charge distribution. Thus, within the dipole approximation we have ${\bf p}=m{\dot {\bf r}}$ in the multipolar-gauge, so that ``matter" is purely mechanical. On the other-hand, ``light" is defined using ${\bf D}_{\rm T} = {\bf E}_{\rm T} + {\bf P}_{\rm T}$ where ${\bf P}_{\rm T}={\bf E}_{\rm L}$ except inside the charge distribution. In the dipole approximation this means ${\bf P}_{\rm T}={\bf E}_{\rm L}$ at all points other than ${\bf 0}$, which is the point dipole's own position. Therefore ``light" within the multipolar-gauge is defined in terms of the total electric field ${\bf E}$ at all points outside the source, which is where the field can be measured. The field ${\bf E}$ is local and it propagates outward causally from the fully localised and bare (purely mechanical) dipole. Thus, in contrast to its Coulomb-gauge counterpart, a multipolar-gauge ``material dipole" is completely bare and fully localised. 

More generally it is possible to define gauges for which ``matter" is defined to be bare matter plus some partial dressing by the electrostatic field, including for example, gauges for which ground state transverse ``virtual" photons are highly suppressed. The ``material" dipole defined relative to this gauge is more localised than its Coulomb-gauge counterpart, but less localised than its fully localised bare multipolar-gauge counterpart. It can be interpreted as providing a definition in which the virtual photons ordinarily dressing the bare dipole have now been absorbed into it's definition. Since virtual processes occur over short timescales the extent to which a dipole is perceived as bare by some pointer device, will depend not only on the length-scales, but also on the time-scales that the pointer is able to resolve.

It is the choice of gauge that directly controls the relevant physical properties of the subsystems by controlling how certain gauge-invariant degrees of freedom are shared between them. We emphasise again that this is a form of vector space relativity that is fully consistent with gauge-invariance, but which cannot be ignored in ultrastrong-coupling regimes. These points are independent of whether or not a truncation happens to have been performed, breaking the gauge-invariance of the theory. Gauge non-invariance is a separate aspect that can only result from approximations and that can always be avoided, at least in principle, by avoiding the offending approximations.

\section*{Example of a double-well dipole}\label{s4}

In seeking to determine the optimality of different material truncations, recently, two-level models have been compared in highly anharmonic regimes. Here, more than merely being a well-defined mathematical procedure, one expects the two-level truncation to offer a generally robust approximation of the non-truncated theory \cite{de_bernardis_breakdown_2018,stefano_resolution_2019,roth_optimal_2019}. Despite this, several much less anharmonic regimes are of importance for experiments. Let us nevertheless consider a highly anharmonic double-well dipole as in Ref. \cite{stefano_resolution_2019} and compare the different two-level model spectra with the exact (unique) energy spectrum of the non-truncated theory. We consider the resonant case $\delta=\omega/\omega_m=1$ together with a high anharmonicity $\mu=(\omega_{21}-\omega_m)/\omega_m$ of $\mu=70$, as considered in the Supplementary Material of Ref. \cite{stefano_resolution_2019}. We compare the unique spectrum of $H_\alpha$,  with the different approximations given by the spectra of the multipolar-gauge QRM $H_1^2$, the standard Coulomb-gauge QRM $H_0^2$, and the new Coulomb-gauge two-level model ${\tilde H}_0^2$. Note that all multipolar models $h_1^2(\alpha)$ possess the same spectrum as $H_1^2$ because they all belong to the same unitary equivalence class (i.e.~they are Bloch sphere rotations).

\begin{figure}[t]
\begin{minipage}{\columnwidth}
\begin{center}
\hspace*{-1.5mm}\includegraphics[scale=0.38]{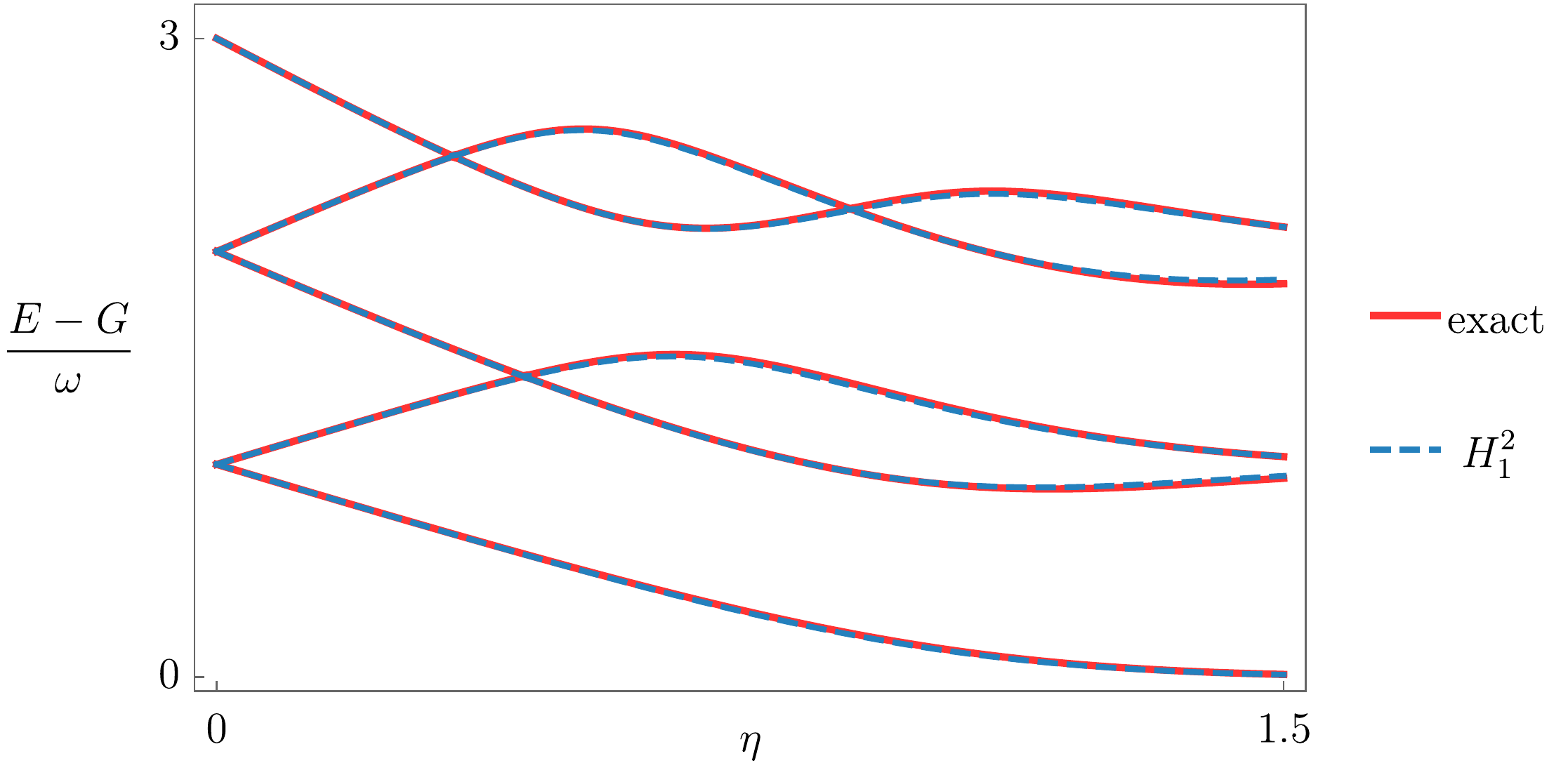}
\caption{The transition spectrum of the multipolar-gauge QRM $H_1^2$ (blue, dashed) is compared with the exact (gauge-invariant) transition spectrum (red, solid), assuming a material anharmonicity of $\mu=70$ and resonance $\delta=1$. The multipolar gauge QRM is {\em generally} accurate in this regime, in the sense that one must go to very high levels before discrepancies with the exact spectrum are found. This graph is the same as Fig.~S2(a) in Ref.~\cite{stefano_resolution_2019}. All multipolar models $h_1^2(\alpha)$ possess the same spectrum as $H_1^2$ because they all belong to the same unitary equivalence class, i.e., they are Bloch sphere rotations.}\label{f1}
\end{center}
\end{minipage}
\end{figure}
\begin{figure}[t]
\begin{minipage}{\columnwidth}
\begin{center}
\hspace*{-1.5mm}\includegraphics[scale=0.38]{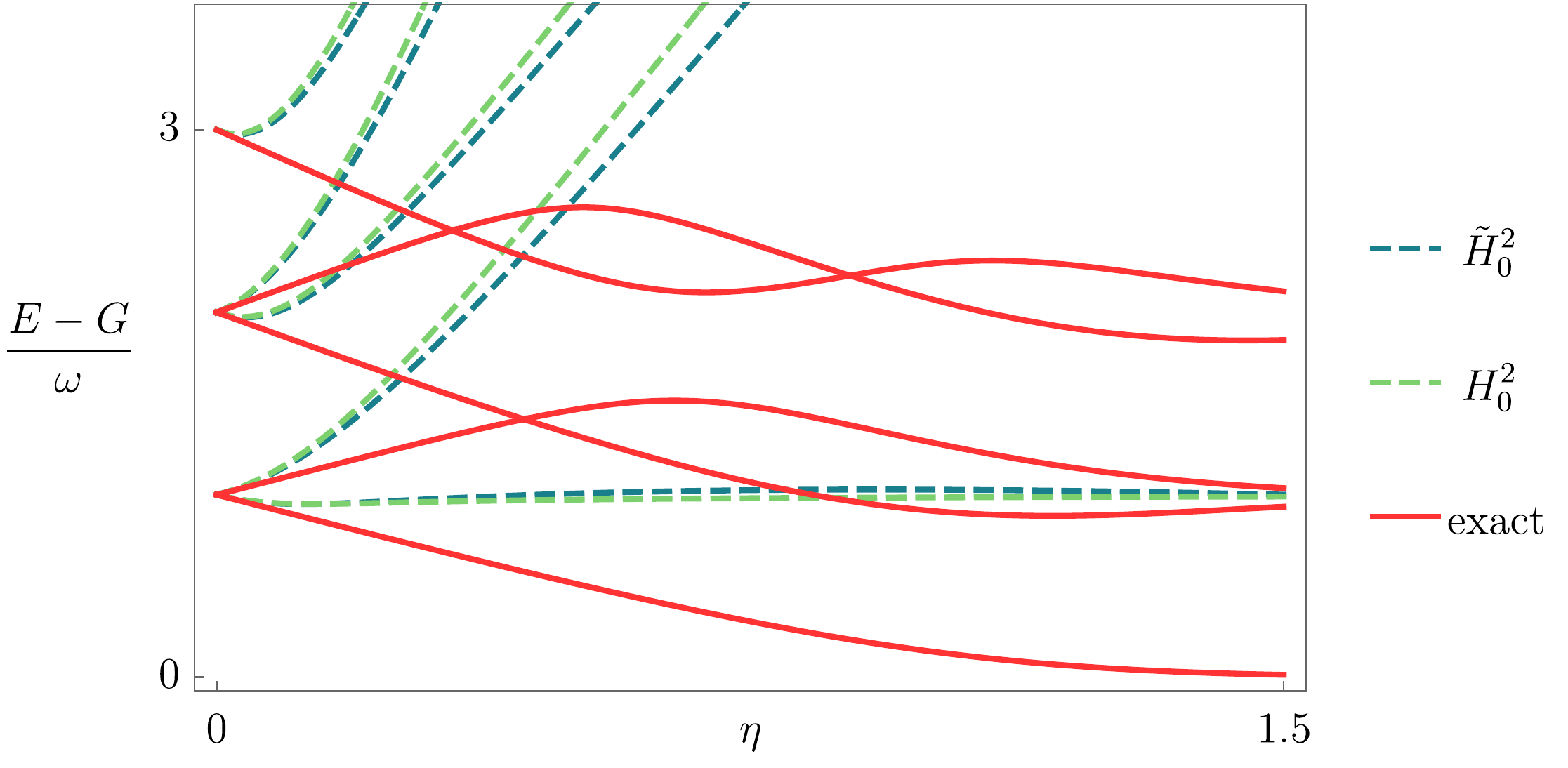}
\caption{The transition spectra of {the standard Coulomb-gauge QRM} $H_0^2$ (light-green, dashed) and {the modified Coulomb-gauge model} ${\tilde H}_0^2$ (dark-green, dashed) are compared with the exact transition spectrum (red, solid) assuming the same parameters as in Fig. \ref{f1}. Both Coulomb-gauge two-level models are generally inaccurate, and are qualitatively very similar.}\label{f2}
\end{center}
\end{minipage}
\end{figure}

The multipolar-gauge QRM $H_1^2$ is very accurate for predicting transition spectra in this regime (Fig.~\ref{f1}) while the Coulomb-gauge QRM $H_0^2$ is very inaccurate for strong enough couplings (Fig.~\ref{f2}). 
An important question however, is whether the spectrum of ${\tilde H}_0^2$  is the same as that of $H_1^2$ as claimed in Ref.~\cite{stefano_resolution_2019}. Were this the case {then} the specific problem of obtaining for gauges $\alpha\neq 1$ {\em generally} accurate two-level models would be resolved. By {\em generally} we mean two-level models that are accurate for more than just the lowest two total-system eigenenergies. However, in regimes where one expects to be able to obtain generally accurate two-level models the spectrum of the new model ${\tilde H}_0^2$ is inaccurate, and is similar to that of the standard model $H_0^2$ as shown in Fig.~\ref{f2}. We remark that this demonstrates very directly that ${\cal T}_{\alpha\alpha'}\neq {\cal G}_{\alpha\alpha'}$.

\section*{Phase-invariance with respect to truncated position}

Supplementary Note 1 of Ref.~\cite{stefano_resolution_2019} provides an alternative derivation of the multipolar equivalence class $\{h_1^2(\alpha')\}$ as resulting from the imposition of a phase-invariance principle defined using the truncated position operator $x_P:=PxP$. We show that this principle can be applied in any gauge $\alpha$ and that it yields the equivalence class $\{h_\alpha^2(\alpha')\}$. The particular class $\{h_1^2(\alpha')\}$ results from invoking this principle after choosing to truncate the theory within the multipolar-gauge. A unitary transformation within a truncated space is mathematically incapable of transforming between distinct minimal-coupling prescriptions, which is the definition of a gauge transformation. Therefore distinct equivalence classes cannot be connected by a rotation such as ${\cal T}_{\alpha\alpha}$, which instead yields equivalent models that all satisfy $x_P$-phase-invariance. Furthermore, as we have shown, the transformation ${\cal T}_{\alpha\alpha}$ is not approximately equal to a truncated gauge transformation such as ${\cal G}_{\alpha\alpha'}$ or $PR_{\alpha\alpha'}$.

The best that can be hoped for is that the $x_P$-phase-invariance principle somehow provides an argument to favour initial truncation within a particular-gauge, such as the multipolar-gauge. However, the multipolar-gauge truncation has been found to become sub-optimal in strong-coupling regimes when more radiation modes are considered even though truncation can remain generally accurate in other gauges \cite{roth_optimal_2019}. This fact would appear to discount the possibility of obtaining a truly general argument for preferring a particular truncation, even if we insist that the truncation must remain generally accurate, rather than only requiring it to be capable of accurately reproducing certain properties of interest. Therefore, a truncated phase-invariance principle cannot ``resolve" gauge non-invariance due to truncation in that the principle can be applied in any gauge after truncation, it cannot result in transformations between different gauges, and it does not tell us that any one truncation will be universally optimal.

There is a simple explanation for why the multipolar-gauge truncation generally performs much better than the Coulomb-gauge truncation when considering a single radiation mode, this being that unlike the Coulomb-gauge coupling the multipolar-coupling does not scale with material transition frequencies \cite{de_bernardis_breakdown_2018}. Whether or not, in addition to this explanation, the localisation properties of the multipolar-gauge material canonical momentum and vector potential can be used to give an alternative explanation, remains to be seen.

We now describe the $x_P$-phase-invariance principle. We remark that an equivalent description can be given that uses an analogy with lattice gauge-field theory. This analogy has been described recently in Ref.~\cite{savasta_gauge_2020} and is reviewed in the Appendix. We begin by noting that the mechanical energy ${\cal H}_{m}(A_\alpha)$ in Eq.~(\ref{mechen1}) satisfies local phase-invariance (gauge-invariance) in the form
\begin{align}\label{pic2b}
\bra{\psi}{\cal H}_{m}(A_\alpha)\ket{\psi} = \bra{\psi'}{\cal H}_{m}(A'_\alpha)\ket{\psi'}
\end{align}
where $\ket{\psi'}=e^{iq\chi}\ket{\psi}$ and $A'_\alpha=A_\alpha+\nabla \chi$. In particular, the equality $\bra{\psi_\alpha}{\cal H}_{m}(A_\alpha)\ket{\psi_\alpha} = \bra{\psi_{\alpha'}}{\cal H}_{m}(A_{\alpha'})\ket{\psi_{\alpha'}}$ in which $\ket{\psi_{\alpha'}}=R_{\alpha\alpha'}\ket{\psi_\alpha}$, expresses gauge-invariance within the $\alpha$-gauge framework and is a special case of Eq.~(\ref{pic2b}) obtained by letting $\chi = \chi_{\alpha'}-\chi_\alpha$.

To define the class $\{h_1^2(\alpha)\}$, the gauge-fixing transformation $R_{1\alpha}$ was replaced with ${\cal T}_{1\alpha}$ in Eq.~(\ref{halpm2}) and the multipolar-gauge mechanical energy ${\cal H}_m(A_1)=H_m$ in Eq.~(\ref{bareen}) was replaced with its projection $P{\cal H}_m(A_1)P$. More generally however, Eqs.~(\ref{halpm2}) and (\ref{bareen}) are special cases of Eqs.~(\ref{mechg}) and (\ref{mechen1}) respectively. 
If we replace $R_{\alpha\alpha'}$ with ${\cal T}_{\alpha\alpha'}$ and ${\cal H}_{m,\alpha}(A)$ with ${\cal H}^2_{m}(A_\alpha):=P{\cal H}_{m}(A_\alpha)P$ on the right-hand-side of Eq.~(\ref{mechg}), then we obtain a truncated $\alpha'$-dependent mechanical energy analogous to that in Eq.~(\ref{mechg});
\begin{align}\label{tpi1}
{\mathscr H}^2_{m,\alpha}(A_{\alpha'}):={\cal T}_{\alpha\alpha'}{\cal H}^2_{m}(A_\alpha){\cal T}_{\alpha\alpha'}^\dagger
\end{align}
expressed as a transformation of the mechanical energy after two-level truncation in the $\alpha$-gauge. Note that the functional dependence of ${\mathscr H}^2_{m,\alpha}$ on $A_{\alpha'}$ is acquired via the equality ${\cal T}_{\alpha\alpha'} =e^{id\sigma^x(A_{\alpha'}-A_\alpha)}$. The energy ${\mathscr H}^2_{m,\alpha}$ satisfies a form of phase-invariance analogous to Eq.~(\ref{pic2b}) but defined with respect to the truncated position operator $x_P:=PxP$. Such a phase transformation is
\begin{align}\label{phas}
U_{x_P}=e^{iq\chi(x_P)} = e^{i\beta}e^{iq\Lambda x_P}=e^{i\beta} e^{id\Lambda \sigma^x} 
\end{align}
where $\beta$ and $\Lambda$ are constants depending on the choice of function $\chi$. The global phase $e^{i\beta}$ can be ignored. Letting $\ket{\psi_2}=P\ket{\psi}$ denote an arbitrary truncated state we have
\begin{align}\label{tpi2}
&\bra{\psi_2}{\mathscr H}^2_{m,\alpha}(A_{\alpha'})\ket{\psi_2} = \bra{{\psi_2}'}{\mathscr H}^2_{m,\alpha}(A'_{\alpha'})\ket{{\psi_2}'}
\end{align}
where $A'_{\alpha'}=A_{\alpha'}+\partial_{x_P}\chi(x_P)=A_{\alpha'}+\Lambda$ and $\ket{{\psi_2}'}=U_{x_P}\ket{\psi_2}$. 
Thus, we see that ${\mathscr H}^2_{m,\alpha}(A_{\alpha'})$ is the mechanical energy of the $\alpha'$-``gauge" where by ``gauge"-invariance we mean $x_P$-phase-invariance within the $\alpha$-gauge truncated mechanical energy. Subsequently, a ``gauge" transformation of $A_{\alpha'}$ under this principle is $A_{\alpha'}'=A_{\alpha'}+\Lambda$.

To obtain the complete $\alpha'$-dependent Hamiltonian one adds the transverse electromagnetic energy, ${\cal H}_{\rm ph,\alpha'}$, defined in Eq.~(\ref{photen1}) to the mechanical energy, which gives the total energy. Following the standard method of obtaining two-level models we note that $E_{\rm T} =- \Pi-P_{\rm T\alpha'}=-\Pi -\alpha' d \sigma^x/v$ is the transverse electric field within the two-level truncation, and we thereby define the transverse electromagnetic energy ${\mathscr H}_{\rm ph,\alpha'} = {v}(E_T^2+\omega^2A)/2$ within the truncation as
\begin{align}\label{phtf}
{\mathscr H}^2_{{\rm ph},\alpha'} &:= {v\over 2}\left[\left(\Pi+{\alpha' d \sigma^x\over v}\right)^2+\omega^2 A^2\right] \nonumber \\ &= {\cal T}_{\alpha\alpha'}{\mathscr H}^2_{{\rm ph},\alpha}{\cal T}_{\alpha\alpha'}^\dagger = {\cal T}_{0\alpha'}H_{\rm ph}{\cal T}_{0\alpha'}^\dagger.
\end{align} 
The second equality in Eq.~(\ref{phtf}) follows from the fact that unlike when acting on $p$, the transformation ${\cal T}_{\alpha\alpha'}$ has the same effect as a gauge-transformation when acting on $\Pi$, because material truncation does not alter the algebra of photonic operators. Combining Eqs.~(\ref{tpi1}) and (\ref{phtf}) we may now define the full $\alpha'$-dependent two-level model $h_\alpha^2(\alpha')$ as the total energy 
\begin{align}
h_\alpha^2(\alpha') = {\mathscr H}^2_{m,\alpha}(A_{\alpha'}) + {\mathscr H}^2_{{\rm ph},\alpha'} = {\cal T}_{\alpha\alpha'}H_\alpha^2 {\cal T}_{\alpha\alpha'}^\dagger.
\end{align}
Thus, the equivalence class $\{h_\alpha^2(\alpha')\}$ can be obtained as the class of Hamiltonians satisfying $x_P$-phase-invariance after truncation within the $\alpha$-gauge. It is not a coincidence that the class obtained in Ref.~\cite{stefano_resolution_2019} is the multipolar-gauge equivalence class $\{h_1^2(\alpha)\}$, which results if the $x_P$-phase-invariance principle is applied within the multipolar-gauge. Such application has the appearance of an application to the free theory only due to approximations implying $A_1 \approx 0$ so that $p-qA_1\approx p$ and therefore that ${\cal H}_m(A_\alpha) \equiv R_{1\alpha}{\cal H}_m(A_1)R_{1\alpha}^\dagger \approx R_{1\alpha} H_m R_{1\alpha}^\dagger$. Without said approximations it is impossible to generate the mechanical energy of the interacting theory by unitary transformation of the bare Hamiltonian $H_m$.

\section*{Discussion}

Fundamental cavity and circuit QED theories begin from first principles and are constructed so as to yield the correct Maxwell equations (or lumped-element circuit approximations in the case of circuit QED). This theory defines what is meant by gauge-freedom and the material systems encountered therein are necessarily infinite-dimensional. A truncated theory is an approximation to this. Fundamentally, quantum subsystems are gauge-relative, meaning that they represent different physical degrees of freedom in each gauge. As a result of this, a truncation of the ``matter" quantum subsystem in the gauge $\alpha$ constitutes a different procedure physically, to truncation in the gauge $\alpha'\neq \alpha$. Since truncation alters the algebra of material operators the truncated theories derived in different gauges are not equivalent.

\begin{figure}
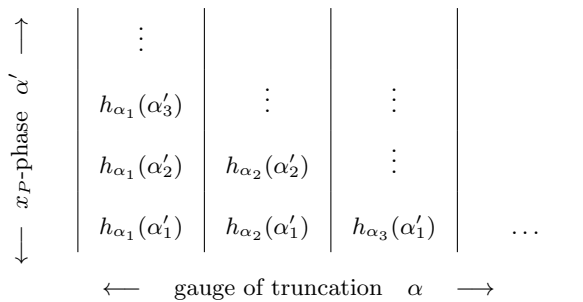

\setlength{\tabcolsep}{8pt} 
\renewcommand{\arraystretch}{2.2}
\hspace*{-4mm}\begin{tabular}{ c|c|c|c|c| } 
 {\multirow{2}{*}{\rotatebox{90}{$\longleftarrow$~~$x_P$-phase~~$\alpha'$~~$\longrightarrow$~}~~~}} & $\vdots$ &  &  & \\ 
 & $h_{\alpha_1}(\alpha'_3)$ & $\vdots$ & $\vdots$ &  \\ 
 & $h_{\alpha_1}(\alpha'_2)$ & $h_{\alpha_2}(\alpha'_2)$ & $\vdots$ &  \\ 
 & $h_{\alpha_1}(\alpha'_1)$ &  $h_{\alpha_2}(\alpha'_1)$ & $h_{\alpha_3}(\alpha'_1)$ & ~~~~$\hdots$~~~~\\
  \multicolumn{5}{c}{~~~$\longleftarrow$ ~~ gauge of truncation~~ $\alpha$ ~~ $\longrightarrow$}\\
\end{tabular}
\caption{The models $h^2_\alpha(\alpha')$ represented as an array with $\alpha$ controlling the gauge in which truncation is performed and with $\alpha'$ controlling the choice of phase for truncated states, defined using $x_P:=PxP$. The columns represent an equivalence class of models and each column belongs to a different gauge, specified by $\alpha$. Without loss of generality, any model from a column can be chosen as the representative of the column. For example, the standard two-level model $H_\alpha^2$ is the particular representative $h_\alpha^2(\alpha)$.}\label{pic}
\end{figure}

It is possible to invoke an $x_P$-phase-invariance principle within the truncated model of any gauge and this yields models equivalent to the standard truncated model of the chosen gauge. It is reasonable to conjecture that in order to define a phase-invariance principle that holds within a truncated space possessing reduced information, one must concurrently reduce the information contained within the phases of states by the same amount. Hence, one may reasonably postulate that a truncated theory should be invariant under phase transformations $e^{iq\chi(x_P)}$ and one may wish to call this postulate a ``gauge"-principle in the truncated state space which is the support of $x_P=PxP$. However, regardless of one's chosen semantic conventions, in the context of cavity and circuit QED systems this principle is distinct from the gauge-principle defined by a more fundamental non-truncated theory, because as we have shown $x_P$-phase invariance holds for every one of the models $H_\alpha^2$, but these models are not equivalent for different gauges $\alpha$.

$x_P$-phase-invariance may constitute a sensible notion of gauge-freedom within already finite-dimensional theories, but it must not be mistaken for gauge-invariance as defined by infinite-dimensional theories that necessarily offer more fundamental descriptions of cavity and circuit QED systems.
Furthermore, that the possibility of imposing a post-truncation phase-invariance principle cannot generally resolve gauge-non-invariance resulting from the truncation, is not only the case when the initial material subsystem is infinite-dimensional. Even if the starting theory consists of an equivalence class $\{H\}$ in which matter is already finite-dimensional, then the truncation of the different members of this class will not, in general, result in equivalent truncated models. 

Ref.~\cite{stefano_resolution_2019} provides one example of the many classes of models derived from $H_\alpha$ that satisfy $x_P$-phase invariance, namely, the class $\{h_1^2(\alpha)\}$. Each model within this class possesses exactly the same predictive power, that is, for fixed $\alpha$ there exist no predictions that can be obtained using any one of the $h^2_\alpha(\alpha')$ that cannot also be obtained using $H_\alpha^2$. In other words, since $\{h_\alpha^2(\alpha')\}$ is an equivalence class, the standard model $H_\alpha^2=h_\alpha^2(\alpha)$ can be selected without loss of generality as the representative of its class. Of course, were this not the case then {\em invariance} under $x_P$-phase transformations would not have been achieved. To the best of our knowledge, non-invariance of models with finite-dimensional matter under the application of $x_P$-phase transformations has never been identified as occurring or as needing to be ``resolved" within light-matter theory. What has been observed, is that the procedure of truncation, meaning the reduction of the dimension of a material system, generally breaks gauge-invariance as defined by the starting theory. 

The two-parameter array of models $\{h^2_\alpha(\alpha')\}$ is depicted in Fig.~\ref{pic}. The parameter $\alpha$ controls the choice of gauge as defined within the fundamental non-truncated theory while the parameter $\alpha'$ parametrises the choice of $x_P$-phase within the truncated $\alpha$-gauge theory. The truncated phase-invariance principle does not by itself dictate that any one equivalence class $\{h^2_\alpha(\alpha')\}$ corresponding to some chosen gauge $\alpha$, will always be more accurate than any other equivalence class for any observable in any parameter regime for any system. 
However, as with any approximation or simplification it is certainly possible to ascertain which gauge's truncated theory will be most accurate for a given observable property of a given system in a given regime. This can always be achieved via direct comparison of the truncated and non-truncated predictions, if not by other means.

We emphasis that we do not intend to suggest that gauge non-invariance renders truncation a useless tool or approximation, or that the truncated models encountered throughout physics and chemistry are useless or too simplistic. It is certainly possible for truncated models to produce accurate and reliable predictions that coincide with those of the gauge-invariant non-truncated theory. At the very least, this is certainly the case in weak-coupling regimes. That the breakdown of strict gauge-invariance can occur due to approximations is hardly surprising. Non-equivalence of different truncated models becomes most apparent within the ultrastrong-coupling regime, but here still, truncated models are capable of remaining accurate for many properties of interest.

\section*{Conclusions}\label{conc}

We have shown that material truncation necessarily ruins the gauge-invariance of a single-mode single-dipole theory. The idea of Ref.~\cite{stefano_resolution_2019} to define two-level model gauge transformations as projections of gauge-fixing transformations does not resolve gauge non-invariance, because it does not produce equivalent models. The Coulomb-gauge model that results is hardly more accurate than the usual Coulomb-gauge QRM in the highly anharmonic regimes considered in Ref.~\cite{stefano_resolution_2019}  where the multipolar gauge QRM is accurate.

The models actually analysed in Ref.~\cite{stefano_resolution_2019} are Bloch-sphere rotations of the multipolar QRM. They all belong to the multipolar gauge and they all possess one and the same spectrum. The Bloch-sphere rotations used to obtain them are mathematically incapable of implementing a gauge change, due to the fundamental modification of the operator algebra that has already been incurred by the truncation. A Bloch-sphere rotation can be used to generate two-level models equivalent to any chosen two-level model. This fact cannot resolve gauge non-invariance due to material truncation. Furthermore, gauge-ambiguities are {broader} than gauge non-invariance due to material truncation.

\section*{Appendix}\label{append}

The Peierls substitution \cite{peierls_zur_1933} can be used to derive the interaction of a magnetic gauge-field with an electron bound in a lattice \cite{graf_electromagnetic_1995}. By drawing an analogy between such a lattice system and a truncated double-well dipole one obtains an alternative way to understand the $x_P$-phase invariance principle for the truncated double-well system \cite{savasta_gauge_2020}. Specifically, if we view this two-level system as analogous to an electron restricted to a single energy band of a two-site lattice, then a substitution analogous to the Peierls substitution can be defined for the two-level system.

The actual Peierls substitution uses a gauge transformation to obtain the Coulomb-gauge mechanical energy from the bare Hamiltonian $H_m$ of the lattice. It is valid under the same electric dipole approximation that for an atomic system would imply that the bare energy $H_m$ coincides with the multipolar-gauge mechanical energy ${\cal H}_m(A_1)$. The analogous ``Peierls substitution" for the truncated double-well dipole, is therefore equivalent to invoking an $x_P$-phase invariance principle within the multipolar-gauge truncated mechanical energy $P{\cal H}_m(A_1)P=PH_mP$. The procedure of defining this ``Peierls substitution" for the two-level system is described below.

We begin by considering a single electron confined within an $N$-site lattice with each site labelled by a position ${\bf R}_l~,l=1,..,N$. The Hamiltonian is \cite{luttinger_effect_1951}
\begin{align}
H_m={{\bf p}^2\over 2m} +V({\bf r})
\end{align}
where $V({\bf r})$ is the periodic potential provided by the lattice. For simplicity we will restrict the electronic excitations to a single-band. The orthonormal electronic energy eigenfunctions are Bloch functions labelled by a single reciprocal lattice index ${\bf k}$;
\begin{align}
H_m \psi_{\bf k}({\bf r}) = E_{\bf k} \psi_{\bf k}({\bf r}).
\end{align}
A localised Wannier function can be defined for each lattice site as
\begin{align}
w_l=w({\bf r}-{\bf R}_l) = {1\over \sqrt{N}} \sum_{\bf k}  \psi_{\bf k}({\bf r}) e^{-i{\bf k}\cdot {\bf R}_l}.
\end{align}
These functions are orthonormal in the sense that $\braket{w_l|w_{l'}}=\delta_{ll'}$ where $\braket{\cdot|\cdot }$ denotes the usual inner-product on $L^2({\mathbb R}^3)$. The matrix representation of $H_m$ in the site basis is denoted $-t$, viz.,
\begin{align}
H_m = -\sum_{l,l'} t_{ll'}\ket{w_l}\bra{w_{l'}}, \qquad t_{ll'} = -\bra{w_l}H_m\ket{w_{l'}}.
\end{align}
Introducing coupling to the transverse vector potential via ${\bf p} \to {\bf p}-q{\bf A}_{\rm T}({\bf r})$ in $H_m$ gives the Coulomb-gauge mechanical energy
\begin{align}
{\cal H}_m[{\bf A}_{\rm T}]={1\over 2m}\left[{\bf p}-q{\bf A}_{\rm T}({\bf r})\right]^2 +V({\bf r}).
\end{align}

The Peierls substitution is a matrix transformation of $t$ that gives an approximation of the corresponding matrix for ${\cal H}_m[{\bf A}_{\rm T}]$. It makes use of the existence of alternative choices of free basis states $\ket{{\bar w}_l}$ and $\ket{w_l}$ that are defined to be related by a PZW transformation in which the vector ${\bf R}_l$ acts as a multipole centre \cite{luttinger_effect_1951};
\begin{align}
\ket{{\bar w}_l} &= e^{-iq\chi_1^l({\bf r})} \ket{w_l}\\
q\chi^l_1({\bf r}) &= -q\int_{{\bf R}_l}^{\bf r} d{\bf s}\cdot {\bf A}_{\rm T}({\bf s}) =-\int d^3 x\,  {\bf P}_{\rm T1}^l({\bf x})\cdot {\bf A}_{\rm T}({\bf x})
\end{align}
where
\begin{align}
{\bf P}_{{\rm T}1}^l({\bf x}) = q\int_0^1 d\lambda \, ({\bf r}-{\bf R}_l)\cdot \delta_{\rm T}({\bf x}-{\bf R}_l-\lambda[{\bf r}-{\bf R}_l])
\end{align}
is the transverse multipolar polarisation connecting the $l$'th lattice site vector ${\bf R}_l$ to the charge position ${\bf r}$.

We can represent the Hamiltonian ${\cal H}_m[{\bf A}_{\rm T}]$ in the basis $\ket{{\bar w}_l}$ as
\begin{align}\label{coullat}
{\cal H}_m[{\bf A}_{\rm T}] = \sum_{l,l'} \bra{{\bar w}_l}{\cal H}_m[{\bf A}_{\rm T}]\ket{{\bar w}_{l'}}\ket{{\bar w}_l}\bra{{\bar w}_{l'}}.
\end{align}
The matrix elements are computed as
\begin{align}\label{matel}
&\bra{{\bar w}_l}{\cal H}_m[{\bf A}_{\rm T}]\ket{{\bar w}_{l'}} \nonumber \\&= \int d^3 r\, e^{iq\chi_1^l({\bf r})}w^*({\bf r}-{\bf R}_l){\cal H}_m[{\bf A}_{\rm T}]e^{-iq\chi_1^{l'}({\bf r})}w({\bf r}-{\bf R}_{l'}) \nonumber \\ &= \int d^3 r\, e^{iq[\chi_1^l({\bf r})-\chi_1^{l'}({\bf r})]}w^*({\bf r}-{\bf R}_l){\cal H}_m[{\bf A}_{1}^{l'}]w({\bf r}-{\bf R}_{l'})
\end{align}
where
\begin{align}
{\bf A}_1^{l'}({\bf r})&={\bf A}_{\rm T}({\bf r})+\nabla \chi_1^{l'}({\bf r}) \nonumber \\
&= -\int_0^1 d\lambda \, \lambda({\bf r}-{\bf R}_{l'})\times {\bf B}({\bf R}_{l'}+\lambda[{\bf r}-{\bf R}_{l'}])
\end{align}
is the multipolar potential referred to the $l'$'th site. Neglecting this potential is nothing but the electric dipole approximation that assumes ${\bf A}_{\rm T}$ does not vary appreciably over the extent of $w({\bf r}-{\bf R}_{l'})$. We then have ${\cal H}_m[{\bf A}_1^{l'}] =H_m$, i.e., the multipolar-gauge mechanical energy is the bare material energy, and Eq.~(\ref{matel}) becomes
\begin{align}\label{matel2}
&\bra{{\bar w}_l}{\cal H}_m[{\bf A}_{\rm T}]\ket{{\bar w}_{l'}} \nonumber \\&= \int d^3 r\, e^{iq[\chi_1^l({\bf r})-\chi_1^{l'}({\bf r})]}w^*({\bf r}-{\bf R}_l)H_mw({\bf r}-{\bf R}_{l'}).
\end{align}
The phase $e^{iq[\chi_1^l({\bf r})-\chi_1^{l'}({\bf r})]}$ can be simplified by noting that the magnetic flux threading the loop $C = {\bf r}\to {\bf R}_l\to {\bf R}_{l'} \to {\bf r}$, is negligible over the extent of the Wannier functions, which are localised at the lattice sites. We therefore have
\begin{align}
 0 = \oint_C d{\bf s} \cdot {\bf A}_{\rm T}({\bf s}) = \chi_1^l({\bf r}) - \chi_1^{l'}({\bf r}) - \int_{{\bf R}_{l'}}^{{\bf R}_l} d{\bf s} \cdot {\bf A}_{\rm T}({\bf s})
\end{align}
and so Eq.~(\ref{matel2}) can be written
\begin{align}\label{matel3}
&\bra{{\bar w}_l}{\cal H}_m[{\bf A}_{\rm T}]\ket{{\bar w}_{l'}} = e^{iq\int_{{\bf R}_{l'}}^{{\bf R}_l} d{\bf s} \cdot {\bf A}_{\rm T}({\bf s})} t_{ll'}.
\end{align}
Substituted into Eq.~(\ref{coullat}) this yields
\begin{align}\label{coullat2}
{\cal H}_m[{\bf A}_{\rm T}] = -\sum_{l,l'} e^{iq\int_{{\bf R}_{l'}}^{{\bf R}_l} d{\bf s} \cdot {\bf A}_{\rm T}({\bf s})} t_{ll'}\ket{{\bar w}_l}\bra{{\bar w}_{l'}}.
\end{align}
Thus, we see that the Coulomb-gauge Hamiltonian can be obtained by making the Peierls substitution $t_{ll'} \to e^{iq\int_{{\bf R}_{l'}}^{{\bf R}_l} d{\bf s} \cdot {\bf A}_{\rm T}({\bf s})} t_{ll'}$ within the free Hamiltonian $H_m = -\sum_{l,l'} t_{ll'}\ket{{\bar w}_l}\bra{{\bar w}_{l'}}$. 

Consider now a truncated one-dimensional double-well two-level system with $PH_mP=\omega_m\sigma^z/2$. Define ``left" and ``right" states $\ket{L}$ and $\ket{R}$ as the antisymmetric and symmetric combinations of the eigenstates $\ket{\epsilon_1}$ and $\ket{\epsilon_0}$ of $\sigma^z$. The Hamiltonian $H_m$ can be written $H_m= -t(\ket{L}\bra{R}+\ket{R}\bra{L})$ where $t=\omega_m/2$. This suggests an analogy in which we view the left and right states as localised Wannier functions of a lattice in which $t_{LL}=t_{RR}=0$ and $t_{RL}=t_{LR}=t$. Define ``lattice sites" as $x_L=\bra{L}x\ket{L}$ and $x_R=\bra{R}x\ket{R}$ where $x$ is the position operator for the double-well system. A ``Peierls substitution" can be defined as $t_{RL} \to e^{iq\int_{x_L}^{x_R} dx A(x)}t_{RL} = :U_{x_R,x_L}t_{RL}$. Assuming that $A(x)=A_0=A$ is the $x$-independent single-mode transverse (i.e. Coulomb-gauge) vector potential, we obtain
\begin{align}
U_{x_R,x_L} =e^{2idA}=({\cal T}_{10})^{2\sigma^x}
\end{align}
where $q(x_R-x_L) = 2q\bra{\epsilon_1}x\ket{\epsilon_0}=2d$. By analogy with Eq.~(\ref{coullat2}) the Coulomb-``gauge" Hamiltonian resulting from the ``Peierls substitution" is defined as
\begin{align}\label{pscoul}
 &-t(U^\dagger_{x_R,x_L}\ket{L}\bra{R}+U_{x_R,x_L} \ket{R}\bra{L})\nonumber \\ 
=&\, {\cal T}_{10}PH_mP{\cal T}_{10}^\dagger  = {\cal T}_{10}{\cal H}^2_m(A_1){\cal T}_{10}^\dagger={\mathscr H}^2_{m,1}(A_0)
\end{align}
where the right-hand-side is the result obtained when the $x_P$-phase-invariance principle is applied to the truncated multipolar-gauge mechanical energy ${\cal H}^2_m(A_1)=PH_mP$. Eq.~(\ref{pscoul}) is unsurprising, because the Peierls substitution for transforming $H_m$ rests on exactly the same (electric-dipole) approximations that ensure the equality ${\cal H}_m(A_1)=H_m$. We note in passing therefore, that assuming a non-uniform transverse potential $A(x)$ within the ``Peierls substitution" does not constitute moving beyond the dipole approximation. The latter is already implicit within the Peierls substitution \cite{luttinger_effect_1951}. As when employing the $x_P$-phase-invariance principle, if we add to ${\mathscr H}_{m,1}^2(A_0)$ the truncated transverse electromagnetic energy, which in the Coulomb-gauge is simply ${\mathscr H}^2_{\rm ph,0}=PH_{\rm ph}P$, then we obtain
\begin{align}
{\mathscr H}^2_{m,1}(A_0) + {\mathscr H}^2_{\rm ph,0} =: h_1^2(0),
\end{align}
which is nothing but the $\alpha'=0$ member of the multipolar-gauge equivalence class $\{h_1^2(\alpha')\}$. 

In summary, we have seen that defining a two-level model ``Peierls substitution"  and applying it to the truncated free theory is equivalent to applying the $x_P$-phase invariance principle within the truncated multipolar-gauge. This is because the Peierls substitution rests on exactly the same approximations that reduce the multipolar-gauge mechanical energy to the bare energy $H_m$ and this is what enables the mechanical energy in any other gauge to be obtained by gauge-transformation of $H_m$. The ``gauge"-invariance of the models that result from truncated ``Peierls substitution" is $x_P$-phase invariance, which does not coincide with gauge-invariance as defined by the non-truncated theory. Being equivalent to invoking $x_P$-phase invariance, the ``Peierls substitution" cannot resolve gauge-non-invariance due to truncation.

\begin{acknowledgments}
This work was supported by the UK Engineering and Physical Sciences Research Council, grant no. EP/N008154/1. 
\end{acknowledgments}

\bibliography{My_Library.bib}

\end{document}